\documentclass[aps,reprint,twocolumn,superscriptaddress,floatfix,showpacs]{revtex4-1}
\usepackage{amsmath}
\usepackage{graphicx,color}
\begin{document}
\title{Radiation emission from braided electrons in interacting wakefields}


\author{Erik Wallin} 
\affiliation{Department of Physics, Chalmers University of Technology, SE--412 96 G\"oteborg, Sweden}
\author{Arkady Gonoskov} 
\affiliation{Department of Physics, Chalmers University of Technology, SE--412 96 G\"oteborg, Sweden}
\affiliation{Institute of Applied Physics, Russian Academy of Sciences, Nizhny Novgorod 603950, Russia}
\affiliation{University of Nizhny Novgorod, Nizhny Novgorod 603950, Russia}
\author{Mattias Marklund}
\affiliation{Department of Physics, Chalmers University of Technology, SE--412 96 G\"oteborg, Sweden}

\begin{abstract}
The radiation emission from electrons wiggling in a laser wakefield acceleration (LWFA) process, being initially considered as a parasitic effect for the electron energy gain, can eventually serve as a novel X-ray source, that could be used for diagnostic purposes. Although several schemes for enhancing the X-ray emission in LWFA has been recently proposed and analyzed, finding an efficient way to use and control these radiation emissions remains an important problem. Based on analytical estimates and 3D particle-in-cell simulations, we here propose and examine a new method utilizing two colliding LWFA patterns with an angle in-between their propagation directions. Varying the angle of collision, the distance of acceleration before the collision and other parameters provide an unprecedented control over the emission parameters. Moreover, we reveal here that for a collision angle of 5$^\circ$, the two wakefields merge into a single LWFA cavity inducing strong and stable collective oscillations between the two trapped electron bunches. This results in an X-ray emission which is strongly peaked, both in the spatial and frequency domain. The basic concept of the proposed scheme may pave a way for using LWFA radiation sources in many important applications, such as phase-contrast radiography.
\end{abstract}

\maketitle

\section*{Introduction}
The radiation emission from electrons, caused by their oscillations during the process of laser wakefield acceleration (LWFA), is a promising mechanism for creating novel X-ray sources with desirable properties, such as compact size, tunable spectral characteristics and short duration \cite{esarey.pre.2002, nemeth.prl.2008, rousse.prl.2004, kneip.nphys.2010, cipiccia.nphys.2011}. These sources may thus find applications in many areas, from diagnostics in medicine and biology to pump-probe measurements at femtosecond time scales. Several concepts were proposed for controlling and enhancing the X-ray emission, including modifying the wave-front \cite{mangles.apl.2009, ma.srep.2016}, as well as using clustering gas jets \cite{chen.srep.2013}, ionization injection \cite{haung.srep.2016} and a curving path of acceleration \cite{chen.arxiv.2015}. Here we present a new concept based on inducing strong oscillations of the accelerated electron beams by colliding two LWFA patterns. We use 3D particle-in-cell (PIC) simulations and analytical estimates to analyze the concept. Our analysis shows that if the angle between the LWFA pattern propagation directions is small enough a new type of nonlinear dynamics arises: both patterns merge into one moving plasma cavity and the two pre-accelerated bunches of electrons start to oscillate around its center, producing intense X-ray emissions with a narrow spectral and spatial structure.

Over the last years LWFA has become a reliable method for creating compact and tunable sources of electrons suitable for producing radiation. Depending on the needed range of photon energies one can use different basic mechanisms, including Thomson scattering, Betatron and Synchrotron emission and Compton scattering~\cite{TaPhuoc2012, Corde2013}. Apart from using external fields of undulators and wigglers \cite{clarke2004science}, X-ray emission can be also caused by laser fields or induced plasma fields, which can be much stronger. For example, collision with a counter propagating laser pulse \cite{Sprangle1992,Catravas2001,Schwoerer2006,TaPhuoc2012, korzhimanov.ufn.2011} and electron oscillations in the laser-produces plasma cavity have been previously studied \cite{kiselev2004, Rousse2004, chen.arxiv.2015}. Here we consider inducing electron oscillation by colliding wakefields and use the angle of collision to get a transition between these cases, focusing on the case of small angle collisions. Radiation from small angle colliding wakefields is also interesting from diagnostic reasons, as well as from proving different properties of radiation. We shall see that this includes a broad frequency range and small angle of emission. 

Wakefield systems can experience filamentation instabilities. It is known from experiments that these filaments can in turn interact, generating what can be considered a precursor to the systematic studies of wakefield interaction found in the literature \cite{Claes-Goran}. Such interaction between two laser wakefields has been considered in a number of works in the literature, mostly in the cases of co-propagating wakefields or head-on collisions. Ren et al. \cite{Ren2000,Ren2001,Ren2002} investigate the nonlinear force between two co-propagating wakefields, and the dynamics due to this interaction. They find that the oscillations due to the nonlinear interaction is typically of frequency $\omega_p$, the electron plasma frequency of the system, and that the two pulses can coalesce due to radiative cooling. Dong et al. \cite{Dong2002} find a strong radiative emission during beam coalescence from initially co-propagating pulses, perhaps not surprisingly indicating the such systems can be a source of high frequency radiation. Wu et al. \cite{Wu2004} discusses the effect of crossed polarisation in a setup similar to the ones above. Sodha and Sharma \cite{Sodha2006} formulate the interaction of co-propagating wakefields in terms of the non-homogeneous dielectric function, also including a discussion on collisions. 
Wen et al. \cite{Wen2010} investigate how charge loading and electron energy is affected by co-propagating wakefields, as a possible means for increasing the wakefield current. Yang et al. \cite{Yang2013} deviate from the limitation of co-propagation and present PIC simulations for two different angles (0.09 and 0.15 rad) of collision between the wakefields, for the purpose of evaluating effects of the angle on charge loading and energy spread. The head-on collision between wakefields was considered by Deng et al. \cite{Deng2014}, using PIC simulations to investigate the wakefield front dynamics as well as the ensuing electron oscillations. The PIC simulations indicate that the collision result in chaotic electron motion and a transverse escape of electrons from the bubble region.    
These references show that there are features of interacting wakefields that can make both a quantitative as well as qualitative difference for the system, as compared to single pulse wakefields. However, a systematic view is lacking, and therefore also a possibility to use the different features of colliding wakefields for possible applications. One such feature is creating a tunable and compact soft X-ray source from such collisions. 

\begin{figure}[ht]
	\includegraphics[width=1.0\columnwidth]{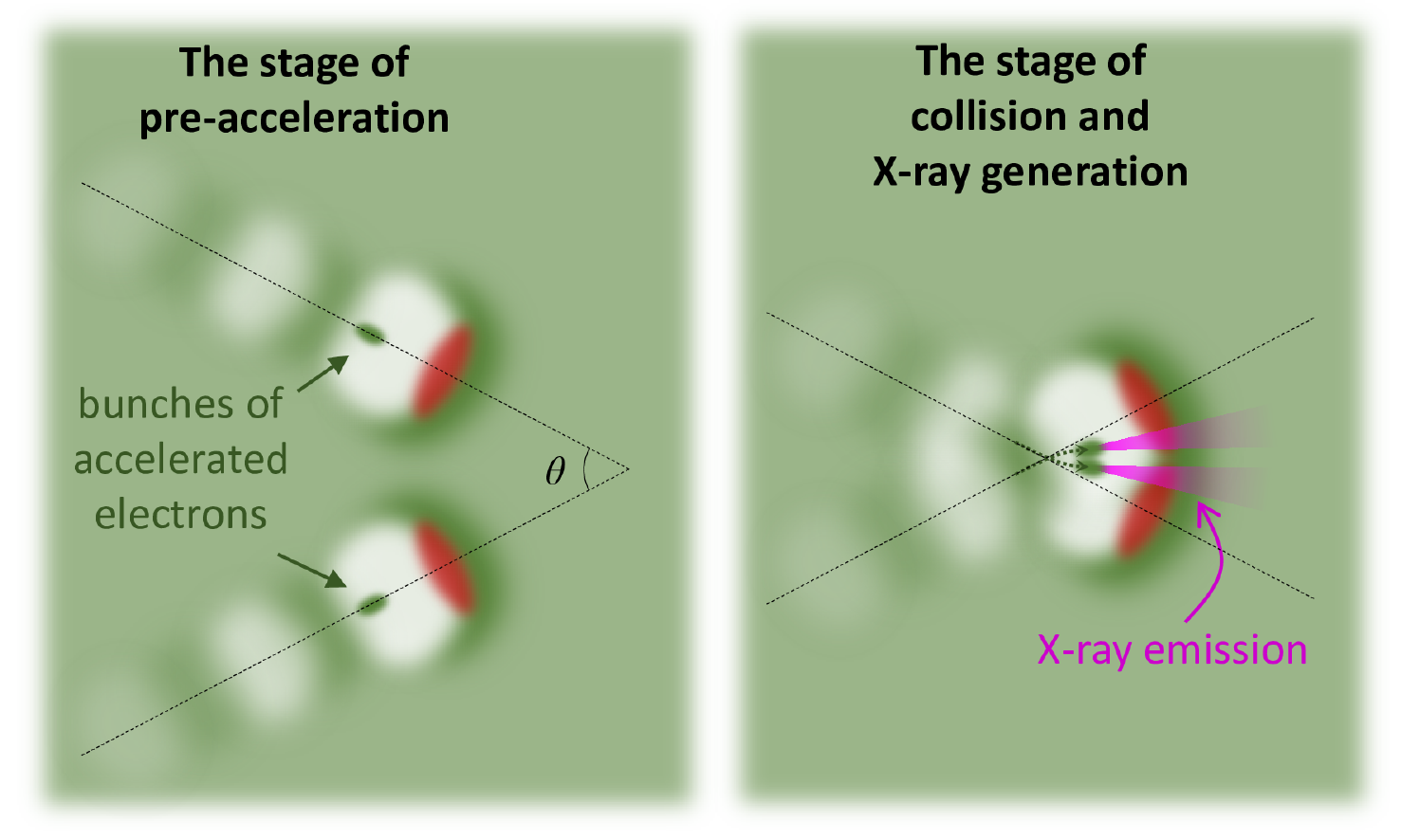}
	\caption{Schematic representation of the collision between two wakefields by an angle $\theta$. For small angle collisions ($\theta < 10^\circ$), the electron bunches will experience nonlinear oscillations in the merged bubble structure, here indicated by the two directions in green.}
	\label{fig:scheme}
\end{figure}

In this paper we do a full parameter scan of collisional angles between wakefields (see Fig.\ \ref{fig:scheme}), in order to control and optimise the electron dynamics for radiation generation. We investigate the corresponding radiative emissions, using both numerical analysis and analytical estimates.
There are two mechanisms that can cause the emitted synchrotron radiation. For larger angles it is mainly the interaction between the electron bunch of one wakefield with the laser pulse of the other. For small angle collisions the regions of electron cavitation behind the laser pulses merge and form a joint large region. The accelerated electrons start to oscillate around the center of this region, which triggers their synchrotron emission.

\begin{figure}[ht]
\centering{\includegraphics[width=1.0\columnwidth]{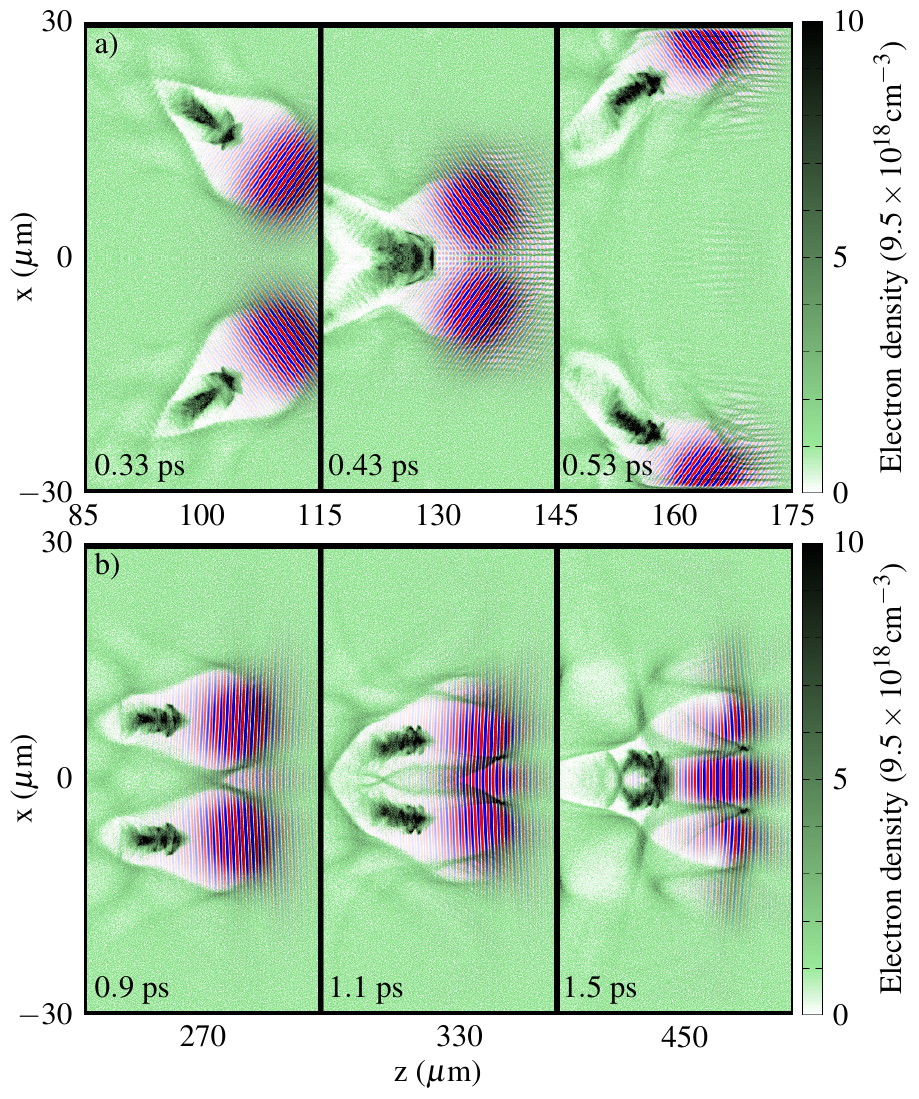}}
  \caption{Collision of two laser wakefields at 70$^{\circ}$ (upper panel) and at 5$^{\circ}$ (lower panel), shown at three different times. The electron density (in the x-z plane, the plane of collision) is plotted in white-green-black in units of the background density and the laser is plotted in blue-red via the y-component of the electric field, with the polarization of the pulse perpendicular to the plane of collision.
}
  \label{fig:deg_coll}
\end{figure}

\subsection*{Governing theory}
We consider the nonlinear regime of LWFA where the laser pulse is of sufficient power to create a region void of electrons. In this ``blowout'' \cite{Rosenzweig1991}, or ``bubble'' \cite{Pukhov2002}, regime the laser pulse in the front is followed by a bubble-shaped region of non-compensated background ions, with a bunch of electrons accelerated by the charge separation in the rear part of the bubble. Colliding two of these fields at angle $\theta$ gives rise to radiation (see Fig.\ \ref{fig:scheme}). The lateral acceleration giving rise to the radiation can be either due to the laser field or the plasma field. In order for the laser-acceleration regime to be prominent the electron bunch should pass through the center of the laser pulse, preferably in a head on collision as the effect of the fields is suppressed for the case of co-propagation. This naturally arises for large angle collisions. For the plasma-acceleration to be efficient the interaction time must be long enough, which is the case for small angle collisions. 

Below we estimate the typical spectra of the radiation in the different regimes, by estimating the \emph{effective magnetic field} $H_{\text{eff}}$ experienced by the electron bunch. From the experimental setup we have the maximum field strength of the pulse given by the dimensionless parameter $a_0$ (relativistic amplitude) and the typical energy of the electrons is given by their Lorentz factor $\gamma$. The typical frequency of the emitted radiation can then be estimated by \cite{LL.V2}
\begin{equation}
  \label{eq:omega_c}
  \omega_c = \frac{3 e H_{\text{eff}}}{2mc}\gamma^2.
\end{equation}

\subsubsection*{Transverse field structure}
For the case of two counter-propagating laser pulses the electron bunch passes through the center of the laser pulse, maximizing the energy and frequency of the radiation. We consider two factors that affect the frequency of the radiation for a collision with an angle $\theta$, a) how the electric- and magnetic fields contribute to $H_{\text{eff}}$ and b) how the distance between the bunch and the pulse center reduces $H_{\text{eff}}$.

For a), we consider the collision between the upper pulse and the lower bunch as seen in Fig. \ref{fig:deg_coll}. The fields are given by
$\mathbf{E} = A \hat{y}$ and $\mathbf{B} = A [ \, \cos(\theta/2) \hat{z} + \sin(\theta/2) \hat{x} \, ]$,
where $A$ is the amplitude of the fields and the velocity of the electrons in the opposite bunch is 
 $ \mathbf{v} \approx c [ \, \cos(\theta/2) \hat{x} + \sin(\theta/2)\hat{z} \, ]$.
Considering the effective magnetic field due to the two terms in the Lorentz force we get
  $\mathbf{F} = q A \hat{y} ( 1 - \cos\theta )$.

For b) we estimate the electron bunch as located $\sim \lambda_p/2$ behind the laser pulse, where $\lambda_p$ is the plasma wavelength. The minimum distance between the bunch and the center of the colliding pulse is then given by
$L_{\text{min}} = \cos(\theta/2) \lambda_p/2$, 
and for a Gaussian shaped pulse of duration $\sim \lambda_p/c$ (FWHM for intensity) the maximal \emph{effective} magnetic field experienced by an electron in the bunch is
\begin{align}
	H_{\text{laser}} = A \big( 1 - \cos \theta \big)
   \exp\left[- \cos^2(\theta/2)  \right]  \ ,
  \label{eq:H_laser}
\end{align}
where $A$ is the pulse's peak amplitude.

\subsubsection*{Longitudinal field structure}
For small angle collisions the radiation is emitted due to the interaction between the electron bunches and the plasma fields. 
The interaction starts when the two bubbles overlap. The plasma is mainly pushed around both laser pulses, creating a larger bubble-like region with the electron bunches offset from the center, as seen in Fig.\ \ref{fig:deg_coll}. The electron streams through the laser region which can be seen, are pushed outwards with time and do not strongly affect our considerations.
The bunches are initially offset a distance $R$ from the center, where $R$ is the radius of a bubble. 
It turns out the effect of one bunch on the other is smaller than the effect of the ions one either bunch, as the bunches are roughly co-propagating and the contribution from the electric and magnetic fields are counteracting.
The effective force on a bunch at radius $r$ will be due to the ions in the sphere of radius $r$. For a test particle in the bunch this is given by
$
F = 4 \pi e^2 n_0 r/3
$. 
The resulting oscillation frequency is $\omega_{\text{osc}} = ({4 \pi e^2 n_0}/{3\gamma m})^{1/2} = \omega_p/\sqrt{3\gamma}$. 
We can thus expect an oscillatory motion of the electron bunches, provided the timescale of the collision is longer than the period time of the oscillation. We can estimate the collision time from the angle of collision and bubble size as
\begin{equation}
  \label{eq:collisionTime}
  T_c = \frac{2R}{c \sin(\theta/2)},
\end{equation}
Furthermore the maximum effective magnetic field due to the plasma field is given by the force at a distance $R$,
\begin{equation}
  \label{eq:H_plasma}
  H_{\text{plasma}} = \tfrac{4}{3} \pi e n_0 R
\end{equation}

\section*{Methods}
\subsection*{Simulation setup}
We perform 3D PIC simulations of the processes using the code \emph{ELMIS3D} \cite{Gonoskov2014a}. A linearly polarized laser pulse with energy $1$J, wavelength $0.81\mu$m, diameter $8 \mu$m and duration $20$ fs (both FWHM in intensity), is sent on a plasma with density $N=9.5 \times 10^{18}$ cm$^{-3}$. Electrons are injected through a density gradient with the plasma having a longitudinal density profile. The density increase linearly from $0$ to $3N$ in $10 \mu$m and then decrease linearly to $N$ in another $10 \mu$m, where it is kept for the reminder of the simulation. The simulation box is $60 \mu m \times 60 \mu m \times 60 \mu m$ on $512\times 128 \times 512$ cells and is co-moving with the pulse. We let the laser wakefield propagate $100 \mu m$ through the plasma after which we clone the pulse and wakefield and rotate them, to collide at different angles, 5$^{\circ}$ and 10$^{\circ}$ to 180$^{\circ}$ in steps of 10$^{\circ}$. Both wakefields are rotated an angle $\theta/2$ where $\theta$ is the collision angle.

The two wakefields are synchronized in time and are initially a transverse distance of $30\mu m$ apart, measured from center to center, and the collisions will thus take place at different longitudinal positions, depending on the angle of collision. An example of this can be seen in Fig.\ \ref{fig:deg_coll} where the process is shown at three different times for the cases of $70^{\circ}$ and $5^{\circ}$ collision angle. The x-z plane (plane of collision) is shown, with the polarization of the pulses in the perpendicular y-direction.
We use a method for determining the high frequency radiation from relativistic particles \cite{wallin2015} where the electrons are considered to be in instantaneous circular motion due to an efficient magnetic field $H_{\text{eff}}$, taking into account the contribution from both the electric- and magnetic fields. We then use a Monte Carlo method to sample from the spectra to emit photons in the direction of propagation of the emitting particle. The recoil of each particle due to emission is calculated as a continuous friction force using the Landau-Lifshitz \cite{LL.V2,QEDPIC} expression,
\begin{align}
  \mathbf{F}_{\text{rad}} &= \frac{2}{3} r_0^2 \Bigg\{ \mathbf{E}\times \mathbf{B} + \frac{1}{c} \bigg[ \mathbf{B} \times (\mathbf{B}\times \mathbf{v} ) + (\mathbf{v} \cdot \mathbf{E}) \mathbf{E} \bigg] \notag \\ 
&- \frac{\gamma^2}{c} \left[ \left( \mathbf{E} + \frac{1}{c} \mathbf{v} \times \mathbf{B}\right)^2 - \left( \frac{\mathbf{E} \cdot \mathbf{v}}{c}\right)^2 \right] \mathbf{v} \Bigg\} ,
\label{eq:RRForce_3D}
\end{align}
where $r_0 = {e^2}/{m_e c^2}$ is the classical electron radius in cgs units.

\section*{Results and Discussion}
\begin{figure}[ht]
  \includegraphics[width=1.0\columnwidth]{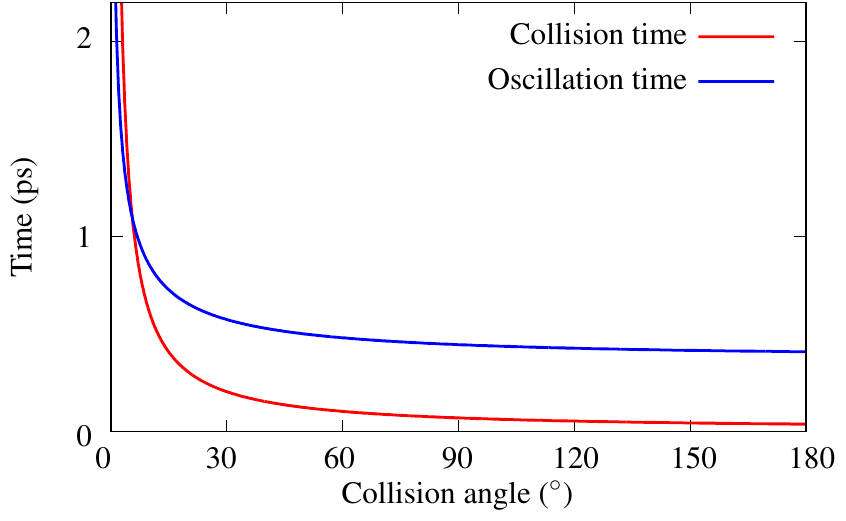}
  \caption{Comparison of the typical collision time and the bunch oscillation period for the simulated wakefield collisions. For small angles the collision time is long enough for oscillations to occur.}
  \label{fig:timeComparison}
\end{figure}

\begin{figure}[ht]
  \includegraphics[width=1.0\columnwidth]{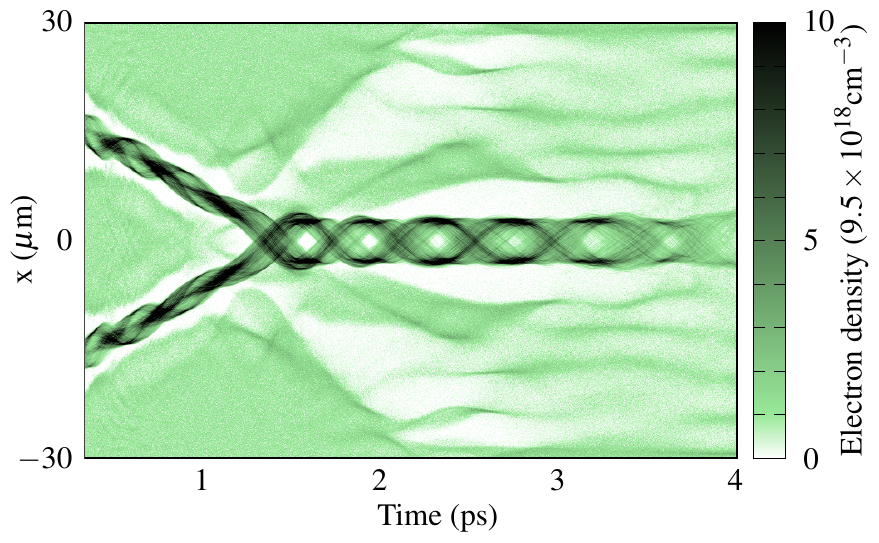}
  \caption{Cross section of the simulation box (electron density) at the bunch position, for the duration of the simulation; showing the bunch oscillations for the case of $5^{\circ}$ collision. The interaction length is $\sim 10$ times longer than the estimate of the typical collision time.}
  \label{fig:braidedElectrons}
\end{figure}

For large angle collisions the wakefields make a single pass, as seen in Fig.\ \ref{fig:deg_coll}, mostly emitting radiation as the bunch interacts with the colliding pulse. However, for small angle collisions we observe the predicted oscillating electron bunches. In Fig.\ \ref{fig:timeComparison} a comparison of the timescales of bunch oscillation and collision time can be seen, in line with the observed oscillations. The oscillatory motion of the bunches can be seen in Fig.\ \ref{fig:braidedElectrons}, showing a cross section of the simulation box for the full extent of the simulation for the $5^{\circ}$ collision. 
The estimated collision time is $\sim 0.5 ps$, however one can see that the bunch oscillations persist longer than so. As the initial wakefields pass, a new larger bubble is formed in which the the electron bunches continue to oscillate. This longer interaction time allows for more energy to be emitted as radiation. It is a remarkable fact that the amplitude of the oscillations in Fig.\ \ref{fig:braidedElectrons} does not decay over time as the electrons radiate. This is a consequence of the relativistic electrons emitting predominantly in the forward direction, so the radiation reaction recoil does not change the direction, only the energy, of the particles.

\begin{figure}[ht]
  \includegraphics[width=1.0\columnwidth]{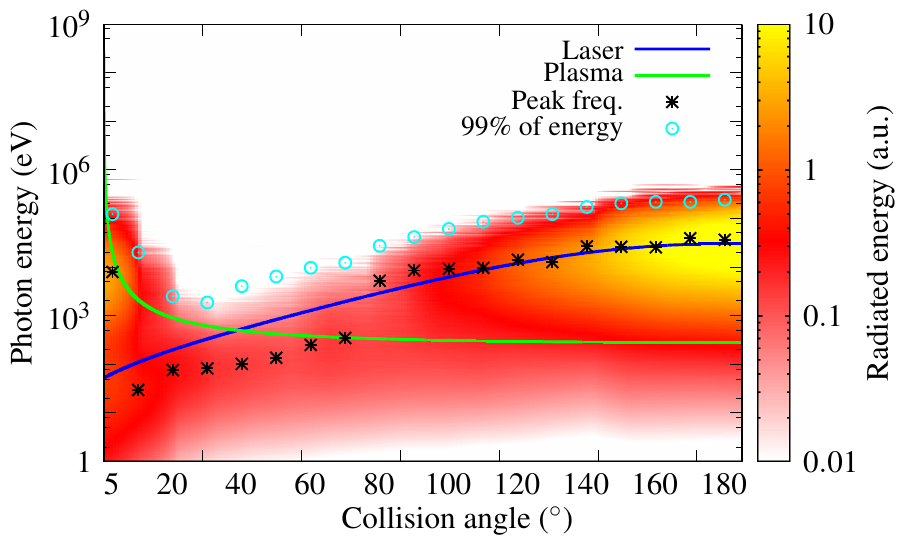}
  \caption{Frequency spectra of the emitted radiation for simulations of collisions for a range of angles. Peak frequencies and frequencies below which 99\% of the energy was emitted are marked. The lines provide estimates of the typical frequency of the emitted radiation due to the laser-(blue) or plasma (green) fields for the simulations.}
  \label{fig:frequencySpectraOverAngle}
\end{figure}

\begin{figure}[ht]
  \includegraphics[width=1.0\columnwidth]{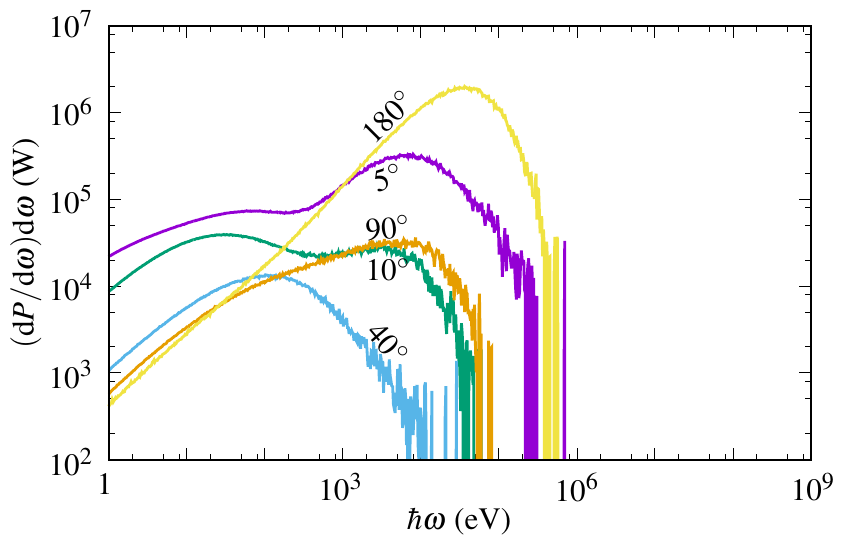}
  \caption{Power spectrum of the emitted radiation for a range of selected collision angles.}
  \label{fig:powerSpectrum}
\end{figure}

\begin{figure}[ht]
	\includegraphics[width=1.0\columnwidth]{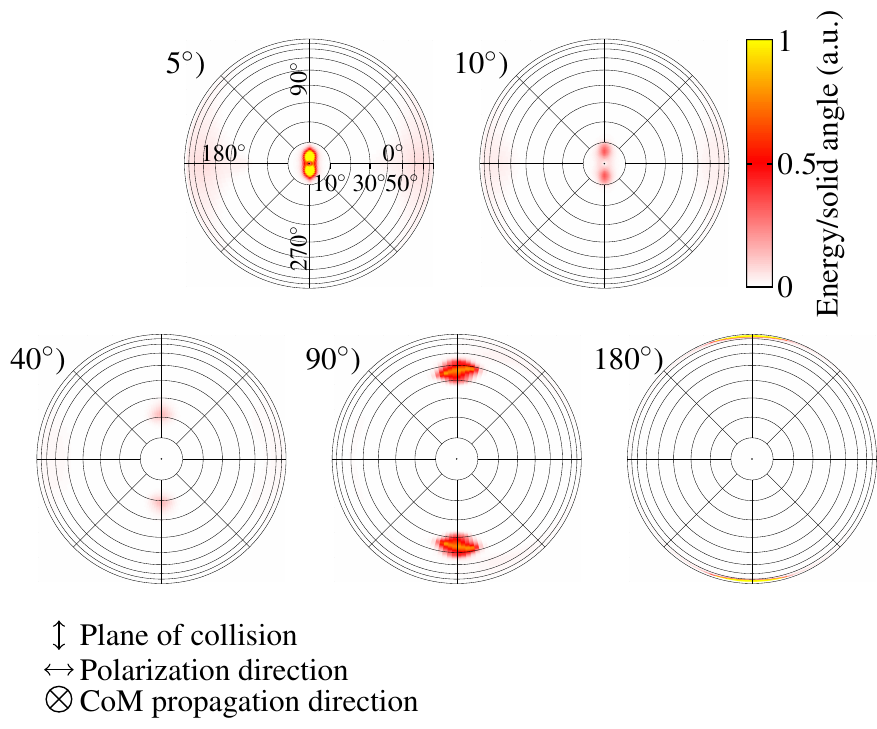}
	\caption{The angular distribution of the emitted photons for the angles 5$^\circ$, 10$^\circ$, 40$^\circ$, 90$^\circ$, and 180$^\circ$. We can see the strong collimation of the radiation for the case of a 5$^\circ$ collision.}
\label{fig:angularPlot}
\end{figure}

\begin{figure}[ht]
  \includegraphics[width=1.0\columnwidth]{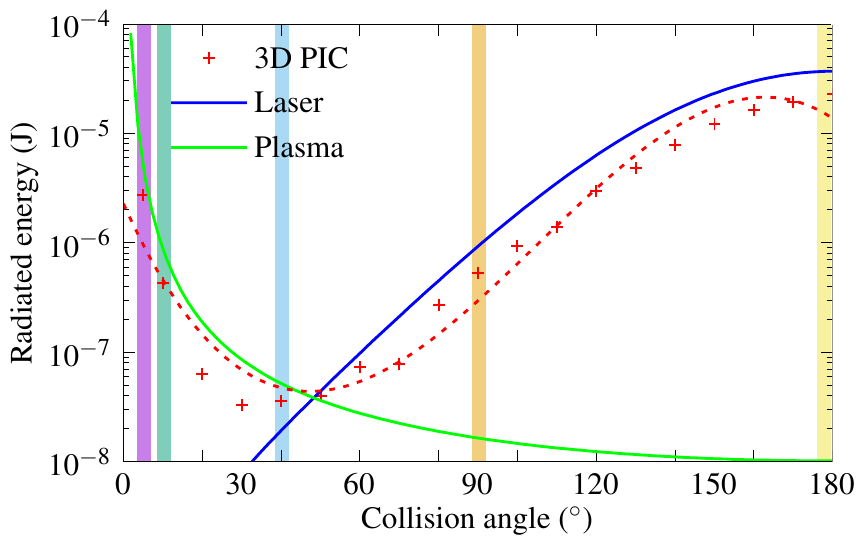}
    \caption{The total emitted radiation as a function of angle for the simulations, with analytical estimates for the two regimes. The marks show the results from 3D PIC simulations, with estimates of the emission due to acceleration via the laser fields (blue) and the plasma fields (green). The dashed line is a polynomial fit to highlight the marks pattern and the vertical background color corresponds to respective power spectra in Fig.\ \ref{fig:powerSpectrum}.}
  \label{fig:emittedEnergyOverCollisionAngle}
\end{figure}

The spectra of the emitted radiation as a function of the collision angle can be seen in Fig.\ \ref{fig:frequencySpectraOverAngle}. Here one can see the two regimes of radiation, with peaks in frequency and total energy of the radiation provided by the collision angles of $\theta \sim 0^{\circ}$ and $\theta \sim 180^{\circ}$. The marks represents the maximum frequency of the radiation and the lines are given by the estimates of the typical emitted frequencies according to Eqs. \ref{eq:omega_c}, \ref{eq:H_laser} and \ref{eq:H_plasma}. These agree well with the radiation for the large and small angle collisions. The emitted energy for the different collision angles can be seen in Fig.\ \ref{fig:emittedEnergyOverCollisionAngle}. This is calculated from the start of the collision to a time when the pulses have passed each other and switched position, thus the shorter time the greater the angle of collision. The radiation from the small angle collisions is emitted within a small angle, with some noise at larger angles due to the plasma particles pushed around the laser pulse, which has been excluded from Fig.\ \ref{fig:emittedEnergyOverCollisionAngle}. Angular plots of the emitted radiation can be seen in Fig.\ \ref{fig:angularPlot} for some selected collision angles and Fig.\ \ref{fig:powerSpectrum} shows a power spectrum for the same selected angles. 


The oscillating electrons and the formation of a joint bubble for small angle collisions provide the possibility of generating radiation from a plasma field with a very long interaction time. One could imagine a situation with balance between the gain of energy due to the wakefield and loss of energy due to radiation for the electrons, resulting in a stable conversion of laser energy into X-rays. 

For the described setup the different collisional angles result in different propagation time before the collisions, and thus different particle energy. In the following sections we attempt a more general comparison between the the two regimes.

\subsection*{Comparison between radiation via laser- or plasma fields}
The power of the emitted radiation for an ultra-relativistic electron in a perpendicular field $H$ is \cite{LL.V2}
\begin{equation}
  I = \frac{2e^4 H^2}{3 m^2 c^3} \gamma^2.
\end{equation}
Using the estimate of the typical collision time given in Eq.~(\ref{eq:collisionTime}) we can make an estimation of the total emitted energy as $E_{\text{rad}}=I T_c$. As seen from the simulations, in the small angle regime the bubbles merge and the oscillations continue longer than the typical collision time. In the simulations this was increased by a factor $\sim 10$.

For the case of the plasma field the field strength is given by Eq.~(\ref{eq:H_plasma}). We estimate the size of the bubble from the density and intensity according to \cite{lu2006nonlinear} 
$
  k_p w_0 \approx k_p R \approx 2 \sqrt{a_0} 
$, 
where $k_p$ is the plasma wave number,  $w_0$ is the laser spot size and $R$ is the radius of the bubble. From this we can get $H_{\text{plasma}}$ (\ref{eq:H_plasma}) only depending on the density, as the bubble radius can be expressed as
\begin{equation}
  R = \frac{2 \sqrt{a_0} c}{\omega_p}.
  \label{eq:size2}
\end{equation}
The normalised field amplitude $a_0$ is related to the field strength $A$ through
$
  a_0 = {e |\mathbf{E}|}/{m_e \omega_0 c} = {e A}/{m_e \omega_0 c}  
$
where $\omega_0$ is the frequency of the laser. We can express the efficient magnetic field for the case of laser acceleration as $H_{\text{laser}} \sim a_0 m_e \omega_0 c / e$. 
Using this we can estimate the total emitted energy for each regime, seen in Fig.\ \ref{fig:emittedEnergyOverCollisionAngle}, where we have compensated for the longer interaction time for the small angle collisions.

Furthermore, keeping the particle energy the same for both cases we can compare the field strength for the two regimes, with their ratio given by
\begin{equation}
  \label{eq:Hp/Hl_ratio}
  \frac{H_{\text{plasma}}}{H_{\text{laser}}} = \frac{2 }{3 a_0^{1/2}} \frac{\omega_p}{\omega_0} \ .
\end{equation}

\section*{Conclusions}
In this paper, we propose and examine a new concept for creating a tunable X-ray source based on the small-angle collision of two laser wakefields. The interacting wakefields provide the deflecting forces, acting as a nonlinear undulator, for the bunches of pre-accelerated electrons. As compared to the concepts of head-on collision and stimulated betatron emission in LWFA, the concept provides a unique control over the emission properties. This control is achieved by varying the collision angle, the plasma density, the laser intensity, and the propagation distance before the collision. In such a way, one can create an X-ray source with tunable photon energy, directivity and beam duration, with a multitude of possible applications.

In this paper, we also identified the nontrivial phenomenon of merging two wakefields into a single one in the case of a sufficiently small angle of collision. This phenomenon provides an opportunity for triggering long-lasting oscillations of the accelerated electron bunches, with an oscillation amplitude as large as the merged bubble size. This scenario can be used for creating a highly efficient converter of the laser pulse energy into X-ray radiation.

\section*{Acknowledgements}
This research was supported by the Swedish Research Council Grants \# 2013-4248 and 2012-5644, and the Knut \& Alice Wallenberg Foundation Grant \textit{Plasma based compact ion sources}. The simulations were performed on resources provided by the Swedish National Infrastructure for Computing (SNIC)  at High Performance Computing Center North (HPC2N). 
The authors thank O.\ Lundh and C.-G.~Wahlstr\"om for valuable discussions. 

\bibliography{References}

\begin{thebibliography}{10}

\bibitem{esarey.pre.2002}
E.~Esarey, B.~A. Shadwick, P.~Catravas, and W.~P. Leemans.
\newblock Synchrotron radiation from electron beams in plasma-focusing
  channels.
\newblock {\em Phys. Rev. E}, 65:056505, May 2002.

\bibitem{nemeth.prl.2008}
K\'aroly N\'emeth, Baifei Shen, Yuelin Li, Hairong Shang, Robert Crowell,
  Katherine~C. Harkay, and John~R. Cary.
\newblock Laser-driven coherent betatron oscillation in a laser-wakefield
  cavity.
\newblock {\em Phys. Rev. Lett.}, 100:095002, Mar 2008.

\bibitem{rousse.prl.2004}
Antoine Rousse, Kim~Ta Phuoc, Rahul Shah, Alexander Pukhov, Eric Lefebvre,
  Victor Malka, Sergey Kiselev, Fr\'ederic Burgy, Jean-Philippe Rousseau,
  Donald Umstadter, and Dani\'ele Hulin.
\newblock Production of a kev x-ray beam from synchrotron radiation in
  relativistic laser-plasma interaction.
\newblock {\em Phys. Rev. Lett.}, 93:135005, Sep 2004.

\bibitem{kneip.nphys.2010}
S.~{Kneip}, C.~{McGuffey}, J.~L. {Martins}, S.~F. {Martins}, C.~{Bellei},
  V.~{Chvykov}, F.~{Dollar}, R.~{Fonseca}, C.~{Huntington}, G.~{Kalintchenko},
  A.~{Maksimchuk}, S.~P.~D. {Mangles}, T.~{Matsuoka}, S.~R. {Nagel}, C.~A.~J.
  {Palmer}, J.~{Schreiber}, K.~T. {Phuoc}, A.~G.~R. {Thomas}, V.~{Yanovsky},
  L.~O. {Silva}, K.~{Krushelnick}, and Z.~{Najmudin}.
\newblock {Bright spatially coherent synchrotron X-rays from a table-top
  source}.
\newblock {\em Nature Physics}, 6:980--983, December 2010.

\bibitem{cipiccia.nphys.2011}
S.~{Cipiccia}, M.~R. {Islam}, B.~{Ersfeld}, R.~P. {Shanks}, E.~{Brunetti},
  G.~{Vieux}, X.~{Yang}, R.~C. {Issac}, S.~M. {Wiggins}, G.~H. {Welsh}, M.-P.
  {Anania}, D.~{Maneuski}, R.~{Montgomery}, G.~{Smith}, M.~{Hoek}, D.~J.
  {Hamilton}, N.~R.~C. {Lemos}, D.~{Symes}, P.~P. {Rajeev}, V.~O. {Shea}, J.~M.
  {Dias}, and D.~A. {Jaroszynski}.
\newblock {Gamma-rays from harmonically resonant betatron oscillations in a
  plasma wake}.
\newblock {\em Nature Physics}, 7:867--871, November 2011.

\bibitem{mangles.apl.2009}
S.~P.~D. Mangles, G.~Genoud, S.~Kneip, M.~Burza, K.~Cassou, B.~Cros, N.~P.
  Dover, C.~Kamperidis, Z.~Najmudin, A.~Persson, J.~Schreiber, F.~Wojda, and
  C.-G. Wahlström.
\newblock Controlling the spectrum of x-rays generated in a laser-plasma
  accelerator by tailoring the laser wavefront.
\newblock {\em Applied Physics Letters}, 95(18):181106, 2009.

\bibitem{ma.srep.2016}
Yong Ma, Liming Chen, Dazhang Li, Wenchao Yan, Kai Huang, Min Chen, Zhengming
  Sheng, Kazuhisa Nakajima, Toshiki Tajima, and Jie Zhang.
\newblock Generation of femtosecond gamma-ray bursts stimulated by laser-driven
  hosing evolution.
\newblock {\em Scientific Reports}, 6:30491, jul 2016.

\bibitem{chen.srep.2013}
L.~M. Chen, W.~C. Yan, D.~Z. Li, Z.~D. Hu, L.~Zhang, W.~M. Wang, N.~Hafz, J.~Y.
  Mao, K.~Huang, Y.~Ma, J.~R. Zhao, J.~L. Ma, Y.~T. Li, X.~Lu, Z.~M. Sheng,
  Z.~Y. Wei, J.~Gao, and J.~Zhang.
\newblock Bright betatron x-ray radiation from a laser-driven-clustering gas
  target.
\newblock {\em Scientific Reports}, 3, may 2013.

\bibitem{haung.srep.2016}
K.~Huang, Y.~F. Li, D.~Z. Li, L.~M. Chen, M.~Z. Tao, Y.~Ma, J.~R. Zhao, M.~H.
  Li, M.~Chen, M.~Mirzaie, N.~Hafz, T.~Sokollik, Z.~M. Sheng, and J.~Zhang.
\newblock Resonantly enhanced betatron hard x-rays from ionization injected
  electrons in a laser plasma accelerator.
\newblock {\em Scientific Reports}, 6:27633, jun 2016.

\bibitem{chen.arxiv.2015}
Min Chen, Fei-Yu Li, Ji~Luo, Feng Liu, Zheng-Ming Sheng, and Jie Zhang.
\newblock A palmtop synchrotron-like radiation source.
\newblock {\em arxiv:1503.08311}, 2015.

\bibitem{TaPhuoc2012}
K.~{Ta Phuoc}, S.~Corde, C.~Thaury, V.~Malka, A.~Tafzi, J.~P. Goddet, R.~C.
  Shah, S.~Sebban, and A.~Rousse.
\newblock {All-optical Compton gamma-ray source}.
\newblock {\em Nat. Photonics}, 6(5):308--311, 2012.

\bibitem{Corde2013}
S.~Corde, K.~{Ta Phuoc}, G.~Lambert, R.~Fitour, V.~Malka, a.~Rousse, a.~Beck,
  and E.~Lefebvre.
\newblock {Femtosecond x rays from laser-plasma accelerators}.
\newblock {\em Rev. Mod. Phys.}, 85(1):1--48, jan 2013.

\bibitem{clarke2004science}
J.~A. Clarke.
\newblock {\em The science and technology of undulators and wigglers}.
\newblock Number~4. Oxford University Press on Demand, 2004.

\bibitem{Sprangle1992}
P.~Sprangle, A.~Ting, E.~Esarey, and A.~Fisher.
\newblock {Tunable, short pulse hard x-rays from a compact laser synchrotron
  source}.
\newblock {\em J. Appl. Phys.}, 72(11):5032--5038, 1992.

\bibitem{Catravas2001}
P.~Catravas, E.~Esarey, and W.~P. Leemans.
\newblock {Femtosecond x-rays from Thomson scattering using laser wakefield
  accelerators}.
\newblock {\em Meas. Sci. Technol.}, 12(11):1828--1834, 2001.

\bibitem{Schwoerer2006}
H.~Schwoerer, B.~Liesfeld, H.~P. Schlenvoigt, K.~U. Amthor, and R.~Sauerbrey.
\newblock {Thomson-backscattered X rays from laser-accelerated electrons}.
\newblock {\em Phys. Rev. Lett.}, 96(1):1--4, 2006.

\bibitem{korzhimanov.ufn.2011}
A.~V. Korzhimanov, A.~A. Gonoskov, E.~A. Khazanov, and A.~M. Sergeev.
\newblock Horizons of petawatt laser technology.
\newblock {\em Physics-Uspekhi}, 54(1):9--28, 2011.

\bibitem{kiselev2004}
S.~Kiselev, a.~Pukhov, and I.~Kostyukov.
\newblock {X-ray Generation in Strongly Nonlinear Plasma Waves}.
\newblock {\em Phys. Rev. Lett.}, 93(13):135004, sep 2004.

\bibitem{Rousse2004}
A~Rousse, K~T Phuoc, R~Shah, and A~Pukhov.
\newblock {Production of a keV X-ray beam from synchrotron radiation in
  relativistic laser-plasma interaction}.
\newblock {\em Phys. Rev. Lett.}, 93(13):135005, sep 2004.

\bibitem{Claes-Goran}
Claes-G{\"o}ran Wahlstr{\"o}m.
\newblock {\em private communication}.

\bibitem{Ren2000}
C.~Ren, R.G. Hemker, R.A. Fonseca, B.J. Duda, and W.B. Mori.
\newblock {Mutual attraction of laser beams in plasmas: Braided light}.
\newblock {\em Phys. Rev. Lett.}, 85(10):2124--2127, 2000.

\bibitem{Ren2001}
C.~Ren, B.J. Duda, and W.B. Mori.
\newblock {Braiding of two spiraling laser beams due to plasma wave wakes}.
\newblock {\em Phys. Rev. E}, 64(6):67401, 2001.

\bibitem{Ren2002}
C.~Ren, B.~J. Duda, R.~G. Evans, R.~A. Fonseca, R.~G. Hemker, and W.~B. Mori.
\newblock {On the mutual interaction between laser beams in plasmas}.
\newblock {\em Phys. Plasmas}, 9(5):2354--2363, 2002.

\bibitem{Dong2002}
Q.L. Dong, Z.M. Sheng, and J.~Zhang.
\newblock {Self-focusing and merging of two copropagating laser beams in
  underdense plasma}.
\newblock {\em Phys. Rev. E - Stat. Nonlinear, Soft Matter Phys.}, 66(2):2--5,
  2002.

\bibitem{Wu2004}
H.C. Wu, Z.M. Sheng, and J.~Zhang.
\newblock {Interactive dynamics of two copropagating laser beams in underdehse
  plasmas}.
\newblock {\em Phys. Rev. E - Stat. Nonlinear, Soft Matter Phys.}, 70(2
  2):1--5, 2004.

\bibitem{Sodha2006}
M.~S. Sodha and A.~Sharma.
\newblock {Mutual focusing/defocusing of Gaussian electromagnetic beams in
  collisional plasmas}.
\newblock {\em Phys. Plasmas}, 13(5), 2006.

\bibitem{Wen2010}
M.~Wen, B.~Shen, X.~Zhang, L.~Ji, W.~Wang, J.~Xu, and Y.~Yu.
\newblock {Generation of high charged energetic electrons by using
  multiparallel laser pulses}.
\newblock {\em Phys. Plasmas}, 17(10):1--6, 2010.

\bibitem{Yang2013}
L.~Yang, Z.~Deng, C.T. Zhou, M.Y. Yu, and X.~Wang.
\newblock {High-charge energetic electron bunch generated by intersecting laser
  pulses}.
\newblock {\em Phys. Plasmas}, 20(3):0--6, 2013.

\bibitem{Deng2014}
Z.~G. Deng, L.~Yang, C.~T. Zhou, M.~Y. Yu, H.~P. Ying, and X.~G. Wang.
\newblock {Collision of counterpropagating laser-excited wake bubbles}.
\newblock {\em Phys. Rev. E - Stat. Nonlinear, Soft Matter Phys.}, 89(6):1--5,
  2014.

\bibitem{Rosenzweig1991}
J.~B. Rosenzweig, B.~Breizman, T.~Katsouleas, and J.~J. Su.
\newblock {Acceleration and focusing of electrons in two-dimensional nonlinear
  plasma wake fields}.
\newblock {\em Phys. Rev. A}, 44(10), 1991.

\bibitem{Pukhov2002}
A.~Pukhov and J.~Meyer-ter Vehn.
\newblock {Laser wake field acceleration: The highly non-linear broken-wave
  regime}.
\newblock {\em Appl. Phys. B Lasers Opt.}, 74(4-5):355--361, 2002.

\bibitem{LL.V2}
L.D. Landau and E.M. Lifshitz.
\newblock {The Classical Theory of Fields}.
\newblock {\em Elsevier, Oxford}, 1975.

\bibitem{lu2006nonlinear}
W.~Lu, C.~Huang, M.~Zhou, W.B. Mori, and T.~Katsouleas.
\newblock {Nonlinear theory for relativistic plasma wakefields in the blowout
  regime}.
\newblock {\em Phys. Rev. Lett.}, 96(16):165002, 2006.

\bibitem{Gonoskov2014a}
A.~Gonoskov.
\newblock {\em {Ultra-intense laser-plasma interaction for applied and
  fundamental physics}}.
\newblock PhD thesis, 2014.

\bibitem{wallin2015}
E.~Wallin, A.~Gonoskov, and M.~Marklund.
\newblock {Effects of high energy photon emissions in laser generated
  ultra-relativistic plasmas: real-time synchrotron simulations}.
\newblock {\em Phys. Plasmas}, 22:33117, 2015.

\bibitem{QEDPIC}
A.~Gonoskov, S.~Bastrakov, E.~Efimenko, A.~Ilderton, M.~Marklund, I.~Meyerov,
  A.~Muraviev, A.~Sergeev, I.~Surmin, and E.~Wallin.
\newblock {Extended particle-in-cell schemes for physics in ultrastrong laser
  fields: Review and developments}.
\newblock {\em Phys. Rev. E}, 92(2):023305, aug 2015.

\end{thebibliography}
\bibliographystyle{unsrt}

\section*{Author contribution statement}
E.W.\ performed the numerical simulations. E.W., A.G., and M.M.\ contributed equally to the analysis, theoretical studies, and writing of the manuscript. A.G.\ conceived the original idea.
%
%

\end{document}